\documentclass[12pt]{article}

\newcommand{\be}{\begin{equation}}
\newcommand{\ee}{\end{equation}}

\newcommand{\ba}{\begin{eqnarray}}
\newcommand{\ea}{\end{eqnarray}}

\newcommand {\ep}{ \varepsilon}

\newcommand{\ve}{\vert}

\begin{document}

\hsize36truepc\vsize51truepc
\hoffset=-.4truein\voffset=-0.5truein
\setlength{\textheight}{8.5 in}

\begin{titlepage}
\begin{center}
\hfill \\

\hfill {LPTENS-01-02}
\vskip 0.6 in
{\large {\bf{ TWIST FREE ENERGY IN A SPIN GLASS}	}}
\vskip .6 in

{\bf E. Br\'ezin$^{a)}$}  {\it and} {\bf C. De Dominicis$^{b)}$}
\end{center}
\vskip 5mm
\begin{center}
{$^{a)}$ Laboratoire de Physique Th\'eorique, Ecole Normale
Sup\'erieure}\\ {24 rue Lhomond 75231, Paris Cedex 05,
France}{\footnote{
 Unit\'e Mixte de Recherche 8549 du Centre National de la
Recherche
Scientifique et de l'\'Ecole Normale Sup\'erieure.
 }}\\ {\it{brezin @corto.lpt.ens.fr}}\\
{$^{b)}$ Service de Physique Th\'eorique, CE Saclay,\\ 91191
Gif-sur-Yvette, France}\\
{\it{cirano@spht.saclay.cea.fr}}
\end{center}

 \vskip 0.5 cm
\begin {center}{\bf Abstract}\end {center}
The field theory of a short range spin glass with Gaussian random
interactions, is considered near the upper critical
dimension six. In the glassy phase, replica symmetry breaking is
accompanied with  massless Goldstone modes, generated by
the breaking of reparametrization invariance of a Parisi type solution.
Twisted boundary conditions are thus imposed at two opposite ends of the
system in order to study the size dependence of the twist free energy. A
loop-expansion is performed to first order around a twisted background.
It is found, as expected but it is non trivial, that the theory does
renormalize around such backgrounds, as well as for the bulk. However two
main differences appear, in comparison with simple ferromagnetic
transitions : (i) the loop expansion yields a (negative) anomaly in the size
dependence of the free energy, thereby lifting the lower critical
dimension to a value greater than two given by $d_c =
2-\eta(d_c)$ (ii) the free energy is lowered by twisting the boundary
conditions. This sign may reflect a spontaneous spatial non-uniformity of the
order parameter.

\vskip 14.5pt
\end{titlepage}
\setlength{\baselineskip}{1.5\baselineskip}


\section{Introduction}

Spontaneously broken symmetries are characterized by the
existence of several possible pure states. If one imposes
"twisted" boundary conditions , i.e. different pure states at two ends of
the system, the free energy per unit volume will be slightly greater than
the free energy corresponding to one single pure state over the whole
system.

 For
a simple discrete symmetry, such as the $Z_2$-symmetry of Ising-like
systems, one may consider an (hyper)-cubic system with up spins in the
$z=0$ plane, down spins in the $z=L$ plane and for instance periodic
boundary conditions in the transverse directions $x_1, x_2, \cdots
x_{d-1}$. This will generate an interface in the system centered around
some plane
$z=z_0$ and a cost in free energy
\be \label{Ising}\Delta F=  F_{\uparrow,\downarrow}-
F_{\uparrow,\uparrow} =
\sigma L^{d-1}
\ee in which $\sigma (T)$ is the interfacial tension.
As is well-known  the power $(d-1)$ of $L$ in
(\ref{Ising}) implies that the lower critical dimension of systems with a
discrete symmetry is equal to one, i.e. there is no ordered phase unless
$d$ is greater than one.  At leading order the classical
(mean field) configuration for the order parameter, given the boundary
conditions,  is a kink of hyperbolic tangent shape, interpolating between
up and down spins. The fluctuations are given at one-loop order by the
Fredholm determinant of a one-dimensional Schr\"{o}dinger operator in a
$1/{\cosh^2(z-z_0)}$ potential \cite{BF} which, as is well-known, is
solvable analytically. Every term in the loop expansion for the free
energy about mean field theory, is then proportional to
$L^{d-1}$, and the succesive contributions build up the correct exponent
and amplitude for the interfacial tension $\sigma$.

For  continuum spontaneously broken symmetries,
 the situation about the upper critical dimension is
technically  different.
For an N-vector model one considers for definiteness an order parameter,
which is uniform along the vector
$(1,0,\cdots,0)$ in the
$z=0$ plane, and uniform but rotated by an angle $\theta_0$ in the
plane $z=L$,  i.e. lying along the vector
$(\cos{\theta_0},\sin{\theta_0},0,\cdots,0)$. There again one expects
a cost in free energy
\be \Delta F = \sigma(T,\theta_0) L^{d-2} \ee
in agreement with a lower critical dimension equal to two, and with a
"twist" energy $\sigma(T,\theta_0)$ (or spin stiffness constant)
vanishing as
$\theta_0^2$ for small $\theta_0$, ( the ratio $\sigma/\theta_0^2$ is
the helicity modulus \cite{FBJ}).
If it is quite elementary to verify these statements within mean field
theory, not difficult also to check them in the vicinity of the lower
critical dimension $d_l=2$ through the non-linear sigma model \cite{Ch,
Elka}.   Near the upper critical dimension
$d_u=4$, things are not as simple. The mean field solution is not elementary
and one may fear that the loop expansion might be difficult to handle.
However it turns out  \cite {BDD} that for $L$ large,  the analysis of
fluctuations is simply perturbative and finally explicit. It follows from
this analysis that the massless Goldstone modes give, as expected, an
$L^{d-2}$ behaviour in the twist free energy to all orders in the loop
expansion.

For a spin
glass the nature of the broken symmetry in the low
temperature phase is more difficult to visualize. However within the
replica approach, and Parisi's  ansatz for the mean field solution
\cite{Parisi}, there are indeed "replicon" massless Goldstone modes
\cite{DD} (plus "anomalous" massless modes). The broken symmetry at the
origin of those modes may be related to a {\it{ reparametrization
invariance}} of the action. More specifically the mean field solution
depends, in the continuum limit of Parisi's scheme of replica
symmetry breaking for the Edwards-Anderson model, of two functions
$p(t)$ and
$Q(t)$ in which $t$ is the continuum labelling of the steps of
breaking,
$Q(t)$ the Parisi order parameter and $p(t)$, the continuum
limit of the size of the successive boxes in which the $n$ replicas are
divided. The free energy is not a separate function of $Q(t)$ and
$p(t)$ but depends only of $Q$ as a function of $p$, leading for
instance to the simple "gauge choice"
$p(t)=t$ of  Parisi. The existence of massless modes may be related to
this arbitrariness \cite{Tem}. However one does not see any  physical
"external field" which could be used to tune a given specific gauge
choice.  The situation is thus reminiscent of cases such as
superfluid Helium, in which there is no physical conjugate variable
to the order parameter which one could use to fix its phase. However
if one takes two samples, they have no reason to carry the same
phase, and this phase difference manifest itself in Josephson's
junctions for instance.

In this note we report the result of an analysis, in which one imposes
again two different schemes at two ends of the system. In the $z=0$ plane
we have chosen
 the simple Parisi gauge
\be \label {1} p(t, z=0) = t \ee
whereas in the $z=L$ plane we have imposed
\be \label{2} p(t, z=L) = t + h(t) \ee
in which we assume that $h(t)$ is some given infinitesimal  function,
vanishing with $t$, with support $ 0 \leq t \leq \tilde x$. All calculations
have been performed to lowest order in $h(t)$.
The mean field solution, to lowest order in $h(t)$, provides a linear
interpolation beteween the two end planes, and a free energy which is
proportional to $L^{d-2}$ as for the N-vector  model. At one-loop order,
in dimension $d=6-\epsilon$ one finds after a long calculation, whose
details will be reported elsewhere, a free energy for the twist
(\ref{1},\ref{2})  which is proportional to $L^{d-2} \log L$. Those
logarithms, which are caused here by the absence of a mass gap to the
Goldstone modes, change drastically the situation compared with ordered
states.  They may  be exponentiated in the standard way and yield, to
 first order in
$\epsilon$ a twist free energy which is proportional to
\be \Delta F \simeq  - \tau^{2+\ep} L^{d-2-\ep/3} \int_0^{\tilde x}
dt \ h^2(t) ,\ee
in which $\tau$ measures the temperature below the glassy
transition.
This is, up  to one-loop, the approximation to
\be \Delta F \simeq - (\frac{L}{\xi})^{d-2+\eta} .\ee
 Since $\eta$ is negative, this shows that the lower critical
dimension $d_c$ is larger than two, given by
\be d_c = 2-\eta(d_c) \ee. The sign of the result is a
puzzle on  which we have a few comments at the end.

\section {Mean field theory}

The action for the Edwards-Anderson spin glass is written in terms of an
$n\times n$ matrix $Q_{ab}$, in which $a$ and $b$ are replica indices and
\be S = \int d^d x \{\sum_{ab}  \left(\frac{1}{4}(\nabla Q_{ab}(x))^2
+
\frac{\tau}{2}  Q_{ab}^2 + \frac{u}{12} Q_{ab}^4 \right)+
\frac{w}{6}
\sum_{abc} Q_{ab}  Q_{bc} Q_{ca}\}.\ee

 In a  Parisi replica infinite symmetry breaking scheme, one
divides the
$n= p_0$  replicas into $p_0/p_1$ boxes  of size $p_1$; each box of size
$p_1$ is divided into $p_1/p_2$ boxes  of size $p_2$ , and so
on, ad infinitum. The matrix elements $Q_{ab}$ follow those steps and
are characterized by a correlative infinite sequence $Q_0,
Q_1,\cdots$ .

In the continuum limit, we are thus led to an action which depends on two
spatially varying functions $p(t,z)$ and $Q(t,z)$, in which $0<z<L$ and $
t$ refers to the steps in the symmetry breaking scheme. ( In the $(d-1)$
transverse directions, periodic boundary conditions have been imposed,
and the mean field solution is independent of those tranverse space
variables).  In terms of those functions the action reads
\ba \label {S} S/n &=& \frac{L^{d-1}}{4}\int_0^L dz [\int_0^1
dt\{-\frac{\partial Q}{\partial z}
\frac{\partial }{\partial z} (\dot p Q) + \dot p\
(\frac{\tau }{2} Q^2 + \frac{u}{12} Q^4)
\} \nonumber \\&-& \frac{w}{6}\left(
\int_0^1 dt \dot p(t,z) (p(t,z)Q^3(t,z) + 3 Q^2(t,z) \int_t^1 ds \dot
p(s,z) Q(s,z)\right)]\ea ( $\dot p, \dot Q$ denote derivatives with
respect to
$t$).

In the bulk, with non twisted boundary conditions, this action is
manifestly a function of
$Q(p)$ alone, and not separately of $Q(t)$ and $p(t)$.
Indeed the extrema of this free energy are given as solutions of
\be \label {A} A(t) = \tau Q + \frac{u}{3}Q^3 - \frac{w}{2} \left( pQ^2
+\int_0^tds
\frac{dp}{ds}Q^2(s) + 2Q\int_t^1 ds \frac{dp}{ds}Q(s) \right) = 0 .\ee

and Parisi's solution is
\ba \label {bulk}Q(t) &=& \frac{w}{2u} p(t)\hskip 5mm {\rm {for}}
\hskip 5mm 0<t<x_1\nonumber
\\ Q(t) &=& Q_1 \hskip 5mm {\rm {for}}
\hskip 5mm x_1<t<1 \ea
with the Edwards-Anderson order parameter $Q_1$
defined by
\be\label {12} \tau + uQ_1^2 - wQ_1 =0.\ee

With the twisted boundary conditions we obtain two variational equations
for $Q(t,z)$ and $p(t,z)$, which read
\ba \dot Q A + \frac{1}{4} \frac{\partial}{\partial t} (Q
\frac{\partial ^2Q}{\partial z^2}) &=& 0\nonumber \\
\dot p A + \frac{1}{4} \frac{\partial^2}{\partial z^2}(\dot p Q) +
\frac{1}{4}\ \dot p \frac{\partial^2Q}{\partial z^2} &=& 0\ea
in which $A$ is defined in (\ref{A}).

To lowest order in the imposed twist $h(t)$ one checks easily that the
solution is
\be \label {twist}p(t,z) = \frac{2u}{w}Q(t,z) =  t +\frac{z}{L}h(t)\ee
for
$0<t<\tilde x$ ($\tilde x$ is the end of the support of $h(t)$ and
then the solution is the bulk one for $t>\tilde x$. To second
order  in
$h(t)$ though, the bulk proportionnality of $p$ and $Q$ is lost).

The incremental free energy, which follows from this
twisted solution (compared to the bulk one), comes purely from the kinetic
energy (since the bulk relation $p(t,z) = \frac{2u}{w}Q(t,z)$
still holds). The final mean field result is, at lowest order in $h(t)$,

\ba\label{MF} \Delta F_{twist} = - \lim_{n\to 0}
\frac{1}{n}(Z^n_{\rm{twisted}}-Z^n_{\rm{bulk}}) &=&
-(\frac{w}{2u})^2L^{d-2}\int_0^{t_1} dt h(t) ( h(t) + t\dot
h(t))\nonumber \\
&=&-\frac{1}{2}(\frac{w}{2u})^2L^{d-2}\int_0^{t_1} dt \ h^2(t) .\ea
This result holds above the upper critical dimension $d_u=6$ but,
contrary to ordered states, we shall see that it is modified by
fluctuations. We now proceed to the one-loop computation.
\section {One-loop fluctuations around mean field}

The theory is now  extended in dimension $d= 6-\epsilon$, dimensionally
regularized , and later renormalized.\begin{enumerate}
\item
Consider first the {\it{Replicon}} sector where the fluctuation
matrix may be fully diagonalized ( for a review of fluctuations beyond mean
field see
\cite{DD}). Its continuation to the bulk would write, in the continuum limit,
\be \label{14} F_{loop}^{(R)} = \frac{L^d}{2}
\left[-\int_0^1\frac{dt}{2}\int _t^1
\frac {dk}{k}
\frac{\partial}{\partial k}\int _t^1 \frac {dl}{l}
\frac{\partial}{\partial l}\right]\int \frac{d^dp}{(2\pi)^d}\log \left( p^2
+ \frac{g}{2}(k^2+l^2-2t^2)\right)\ee
where we have used the notation
\be g = \frac{w^2}{2u} .\ee
In (\ref{14}) the first bracket comes from the multiplicity of the
replicon modes with associated eigenvalues $p^2 + \Delta_0(k,l;t)$
as in the argument of the logarithm
\be\label {18} \Delta_0(k,l;t) = \frac{g}{2}(k^2+l^2-2t^2).\ee Under the twist
(\ref{twist}), the above bulk result (\ref{14}) is changed in two ways.
First the argument of the logarithm is to be replaced by
\be \log (q_T^2 -\frac{\partial^2}{\partial z^2} + \Delta_0+
\frac{z}{L} \Delta_1) \ee
with
\be \Delta_1 = g[ kh(k)+lh(l) -2th(t)] \ee
In fact there is also a quadratic term $(z/L)^2 \Delta_2$, that has
been omitted here since at one-loop it cancels through the bulk
subtraction.

Expanding now the logarithm to second order in $\Delta_1$ (in order to collect
the quadratic terms in $h(t)$), we obtain as twist contribution the term
\ba\label{17} &&\frac{L^{d-1}}{4} \int_0^1\frac{dt}{2}\int _t^1
\frac {dk}{k}
\frac{\partial}{\partial k}\int _t^1 \frac {dl}{l}
\frac{\partial}{\partial l}\int
\frac{d^{d-1}q_T}{(2\pi)^{d-1}}\nonumber\\&&\int_0^L dz \frac{z}{L}\int_0^L dz'
\frac{z'}{L}\left( \frac{1}{\pi}\int_{-\infty}^{+\infty} dK
\frac{\sin{Kz}\sin{Kz'}}{q_T^2+K^2 +\Delta_0}\right)^2 \Delta_1^2 .\ea
In this expression boundary conditions at the two end planes $z=0$
and $z=L$ have been taken into account ; indeed the fluctuating part
of the field vanishes at those boundaries, leading to the appropriate
basis $\sin{\pi\frac{mz}{L}}$ with $m=1,2,\cdots$.

Secondly,  the twist (\ref{twist}) changes also the multiplicity
itself. This is taken into account via an identity expressing the
reparametrization invariance under a {\it{z-independent}}
shift $\Delta_0\to\Delta_0+\Delta_1$. As a result the total twist contribution
is then obtained by replacing in (\ref{17}) $\frac{zz'}{L^2}$ by $
\displaystyle \frac{zz'-1/2(z^2+z'^2)}{L^2}$.

Performing the K-integration on the modified (\ref{17}) one obtains
\ba \Delta F_{twist}^R =&&
-\frac{L^{d-3}}{8}(\int_0^1\frac{dt}{2}\int_t^1
\frac {dk}{k}
\frac{\partial}{\partial k}\int _t^1 \frac {dl}{l}
\frac{\partial}{\partial l}\nonumber \\&&\int_0^L dz \int_0^L dz'[
\frac{e^{-M\ve z-z'\ve}-e^{-M(z+z')}}{2M}]^2 \Delta_1^2\ea
in which
\be M = q^2_T + \Delta_0.  \ee

Performing the $z,z'$ and finally $ q_T$ integrations, to gather
poles in $\ep$ and logarithms, one obtains :
\be\Delta F_{twist}^{(R)} = -\frac{1}{12} L^{d-2} S_d \ g^2 \int_0^{\tilde
x} dt\   h^2(t)\  [\frac{1}{\ep} + \log L +\cdots] ,\ee
with
\be S_d = \frac{2}{ (4\pi)^{d/2} \Gamma (d/2)}\ee
It is a strong argument in favour of the consistency of the
calculation to see that the one-loop contribution is, like mean-field,
proportional to the integral $\int dt\  h^2(t)$. Indeed, otherwise the
fluctuations would not be renormalized by a simple change in the
coupling constant. In the intermediate steps this final form is far form
obvious. In particular it involves the unexpected identity
\be\int_0^1 \frac{dt}{2}\int_t^1
\frac {dk}{k}
\frac{\partial}{\partial k}\int _t^1 \frac {dl}{l}
\frac{\partial}{\partial l} \Delta^2_1 (k,l;t) = g^2\int_0^{\tilde x}
dt\ h^2(t)\ee
\item
In the {\it{Longitudinal-Anomalous}} (L-A) sector, the fluctuation matrix can
only be diagonalized by blocks \cite{DD} (with blocks of size $(R+1)\times
(R+1)$,  $R$ being equal to the number of steps of replica symmetry
breaking). In the Parisi limit, in which $R$ goes to infinity, the
L-A contribution to the bulk free energy writes \cite{DD}
\be F_{loop}^{(LA)} = \frac{L^d}{2} [\int_0^1 \frac
{dk}{k}\frac{\partial}{\partial k}] \int \frac{d^d \vec p}{(2\pi)^d}
{\rm{tr}}\log [1 + \frac{1}{p^2+ \Delta_0(k,t;t)} B_k(t,s)]
 \ee

  where,again,the first bracket comes from the new multiplicity of the L-A
modes . Besides we have the $(t,s)$ matrix

\be  B_{k}(t,s) = g ({\rm{Inf}}(t,s))[\Theta(k-s)+k\delta(k-s)+2\Theta(s-k)]ds
\ee

  and, as in (\ref{18})

 \ba      \Delta_ 0&=&g(k^2-t^2)/2 \hskip 1cm        t<k\nonumber \\
                     &=&0    \hskip 3cm                t\geq k
\ea

   The calculation is more involved here, but we proceed as in the replicon
sector, collecting quadratic terms in $h$ when performing the twist transform
as in (\ref{twist})  or in $\Delta_0 \to   \Delta_0+(z/L)\Delta_1$. As above
we keep only terms  {\it{quadratic}} in the propagator
$(q_T^2-d^2/dz^2+\Delta_0)^{-1}$, higher order terms being, at one
loop, ultra-violet convergent.In contrast to the replicon sector (where only
the twist of $\Delta_0$ into $\Delta_0+(z/L)\Delta_1$ contributed) we need here
to take care  of the twists over the matrix elements $B_k{(t,s)}$ and over the
multiplicity. Altogether, complicated expressions rearrange themselves to give

 \be\Delta F_{twist}^{(LA)} = \frac{1}{6} L^{d-2} S_d\  g^2 \int_0^{\tilde
x} dt \  h^2(t)\  [\frac{1}{\ep} + \log L +\cdots] .\ee

   Notice that, at one-loop,  there is no contribution in $\log(\tau)$ i.e.
involving the mass
$wQ_1\simeq\tau$ (as  in (\ref{12})). The reason is that
neither $Q_1$ (i.e. $x_1$), nor $p_0=n$, fluctuate under
reparametrization. Even as a correction term to the contribution in
$(zh/L),(z'h/L)$, it would take to expand the logarithm to third order
before $Q_1$ showing up.
\section {Renormalization and scaling}
Before proceeding, one has to take into account the fact that the
$\frac{u}{12} \sum_{ab} \phi_{ab}^4$ coupling is irrelevant, and
"dangerous" since the fluctuations make it singular below dimension
eight \cite {FS, DD}. Indeed in a pure $\frac{w}{6} {\rm{tr}} \phi^3$ theory,
the one-loop contribution has the effect of replacing $u$ by
\be u \to u + 12 w^4 \int \frac{d^d p}{(2\pi)^d}\  \frac{1}{ p^4(p^2 + 2
\tau)^2} = u + 6 S_d w^4 (2\tau)^{-(1+\ep/2)}.\ee
We thus have for the twist free  energy
\be \Delta F_{twist}  = -\frac{S_d}{288 w^6}
L^{d-2}\tau^{2+\ep}\int_0^{\tilde x} dt\  h^2(t)\ [ 1 - \frac{2}{3} w^2
(\frac{1}{\ep} + \log L) +\cdots]\ee
( a factor $S_d$ has been included in $w^2$).
It is now crucial to verify that the replacement of the coupling constant
and temperature by their renormalized counterpart $w_R, \tau_R$ rids us
of the $1/\ep$  poles. The computation of those renormalizations is
easily done in the paramagnetic phase. It gives \cite {MC}
\ba \tau_R &=& \tau [ 1- \frac{4w^2}{\ep}+ O(w^4)] Z \nonumber \\
w_R^2 &=& w^2 [ 1- \frac{4w^2}{\ep}+ O(w^4)] Z^{3/2} \nonumber \\
Z &=&  [ 1+\frac{2w^2}{3\ep}+ O(w^4)] \ea
from which follows
\be \frac{\tau^2}{w^6} = \frac{\tau_R^2}{w_R^6}[1 + \frac{2w_R^2}{3
\ep} + O(w_R^4) ].\ee
The $1/\ep$ pole is thus exactly cancelled and we end up with
\be \Delta F_{twist}  = -\frac{S_d}{288 w_R^6}
L^{d-2}\tau_R^{2+\ep}\int_0^{\tilde x} dt\  h^2(t)\ [ 1 - \frac{2}{3} w_R^2
 \log L +\cdots].\ee
If we substitute to $w_R$ the fixed point $w^*$ , zero of the $\beta$-
function
\be \beta (w_R) = -\frac{\ep}{2} w_R + w_R^3 + O( w_R^5) \ee
one obtains that, to this order, the result exponentiates to
\be \Delta F_{twist}  = -\frac{S_d}{288 w_R^6}
\tau_R^{2+\ep}L^{d-2-\ep/3}\int_0^{\tilde x} dt \ h^2(t). \ee
Introducing the usual critical exponents $\eta$ and $\nu$, whose
$\ep$-expansions are known to be \cite{MC}
\be \eta = -\frac{1}{3} \ep +O(\ep^2) \hskip 2cm \nu = \frac{1}{2} ( 1 +
\frac{5}{6} \ep + O(\ep^2)),\ee
 one may write to this order
\be \Delta F_{twist} \sim - \tau_R^{\nu(d-2+\eta)} L^{d-2+\eta}.\ee
This is reasonable since it gives the final twist free energy as a
function of $L/\xi$ to a power,   which is the expected scaling form
for the ordinary order-disorder transitions :
\be \Delta F_{twist} \sim - (\frac{L}{\xi})^{d-2+\eta} .\ee
There are two differences though  with ordinary transitions.
\begin{itemize}
\item

First the
power of $L/\xi$ is non-canonical . We thus verify, to the order of
one-loop, an extended form of scaling, appropriate when the soft
transverse modes are not isolated, but being at the bottom of a gapless
band, they are no longer infra-red free : their propagator develops a
({\it{negative}}) anomaly $\eta$. At the lower critical dimension $d_c$, the
twist free energy should vanish and thus $d_c$ should be the solution of
the equation
\be \label{38}d_c = 2-\eta(d_c).\ee

This same answer had been anticipated earlier on the basis of scaling
arguments applied to the null overlap replicon sector
\cite{KT}. Using numerical estimates for $\eta$ \cite{KY},one gets from
(\ref{38})  a value of $d_c$ close
to $2.5$, whereas a self consistent mean-field approach for the twisted
free-energy of two  copies, surprisingly yields \cite{FP} exactly $5/2$.

\item
The sign of this twisted free energy is negative. This will be discussed
below.
\end{itemize}

\section {Discussion}
The calculation of fluctuations around the background of a twisted mean
field solution is renormalizable, as expected, in spite of its non
spatial non-uniformity, as already checked in the simple O(N)-model
\cite {BDD}. However contrary to the simple ferromagnetic transitions, the
influence of this twist on the size dependence manifests itself by a
logarithmic dependence in $L$ at one-loop which exponentiates to a negative
anomaly. This generates an increase of the lower critical dimension.

However another difference with ferromagnetic transitions is the sign
dependence of the twist on the free energy : the twisting leads
here to a decrease in the free energy. This is not in contradiction with
the principles of thermodynamics. A similar situation could occur with
an antiferromagnet, since there as well, the free energy may be lowered by
imposing  external fields of opposite signs at two ends of the sample.
In such a circumstance, one expects that the system would spontaneously
breaks spatial uniformity and develop a space dependence in the "gauge"
choice of the replica symmetry breaking.  Unfortunately we are not aware
of any conjugate field to the gauge choice $p(t,z)$ which could be
imposed to improve the mean field starting point.
Transposing to dynamics the instability to
twisting found here, might be the sign of a space dependence of local time
scales, as discussed in recent simulations of finite
dimensional spin glasses \cite {Leticia}.

{\bf{Acknowledgments}}: We have benefited from stimulating
discussions with J.-P.Bouchaud, L. Cugliandolo, S. Franz, J. Lebowitz, M.
M\'ezard,
 G.Parisi, J. Zinn-Justin.
\end{enumerate}


\vskip 5mm

\begin{thebibliography}{99}

\bibitem {Widom}B. Widom, {\it{J. Chem. Phys. }}{\bf{43}} (1965) 3892.
\bibitem{BF} E. Br\'ezin and Sze Shao Feng,		{\it{	Phys.
Rev.}} {\bf{B29}}
 (1984) 472.
\bibitem{FBJ} M.E.Fisher, M.N. Barber and D.Jasnow,, {\it{Phys. Rev.
}}{\bf{A8}} (1973) 1111.
\bibitem{Ch} S. Chakravarty, {\it{Phy. Rev. Lett. }}{\bf{66}} (1991)
481.
\bibitem{Elka} E. Br\'ezin, E. Korutcheva, T. Jolicoeur and J.
Zinn-Justin ,{\it{Journ. Stat. Phys.  }}{\bf{70}} (1993) 483.
\bibitem {BDD}E. Br\'ezin and C. De Dominicis {\it{Eur. Phys. J, to be
published}}
\bibitem {Parisi} G. Parisi, {\it{Phys. Rev. Lett.
}}{\bf{43}}(1979) 1754 ; {\it{J. Phys.}}{\bf{A13}}(1980) L115,
{\it{ibidem}} 1101,{\it{ibidem}} 1887.
\bibitem{DD} C. De Dominicis, I. Kondor and T. Temesvari,
in {\it{Spin glasses and random fields  }},  World
Scientific Singapore,  (1998) 119, (A. P. Young editor).
\bibitem {FS} D.Fisher and H.Sompolinsky {\it{Phys. Rev Lett.}}{\bf{ 54}}
(1985) 1063
\bibitem{KT} C. De Dominicis, I. Kondor and T. Temesvari,
{\it{Int. J. Mod. Phys.  }} {\bf{B7}} (1993) 984;
D.Carlucci,thesis, Pisa (1997).
\bibitem {MC} O. de Alcantara Bonfim, J. Kirkham, A.J. Mc Kane {\it{ P.
Phys}}{\bf{A14}}(1981) 2391
\bibitem{Tem} T. Temesvari,I. Kondor and C. De Dominicis, {\it{Eur.
Phys. J. }}{\bf{B18}}(2000) 493
\bibitem {KY} N.Kawashima,A.P.Young  {\it{Phys.Rev.}}{\bf{B 53}} (1996)R484
\bibitem {FP} S. Franz, G. Parisi and M. Virasoro, {\it{J. Phys I
(France)}} {\bf{4}}(1994) 657.
\bibitem {Leticia} H. E. Castillo, C. Chamon, L. Cugliandolo,
M. P. Kennett,{\it{Heterogeneous aging in spin glasse}} cond-mat/0112272
\end{thebibliography}
\end{document}